\documentclass{PoS}

\usepackage{epsfig,macros_static,macros_ssm}
\usepackage{amssymb,fontenc,times,mathptmx,graphicx}

\title{The B-meson mass splitting from non-perturbative quenched lattice QCD}

\ShortTitle{The B-meson mass splitting}

\author{A.G.~Grozin$^{ab}$, \speaker{D.~Guazzini}$^c$, P.~Marquard$^d$, H.B.~Meyer$^e$,
          J.H.~Piclum$^{ad}$, R.~Sommer$^c$ and M.~Steinhauser$^d$\\
          $ $\\
          $ $\hfill \includegraphics[scale=0.11]{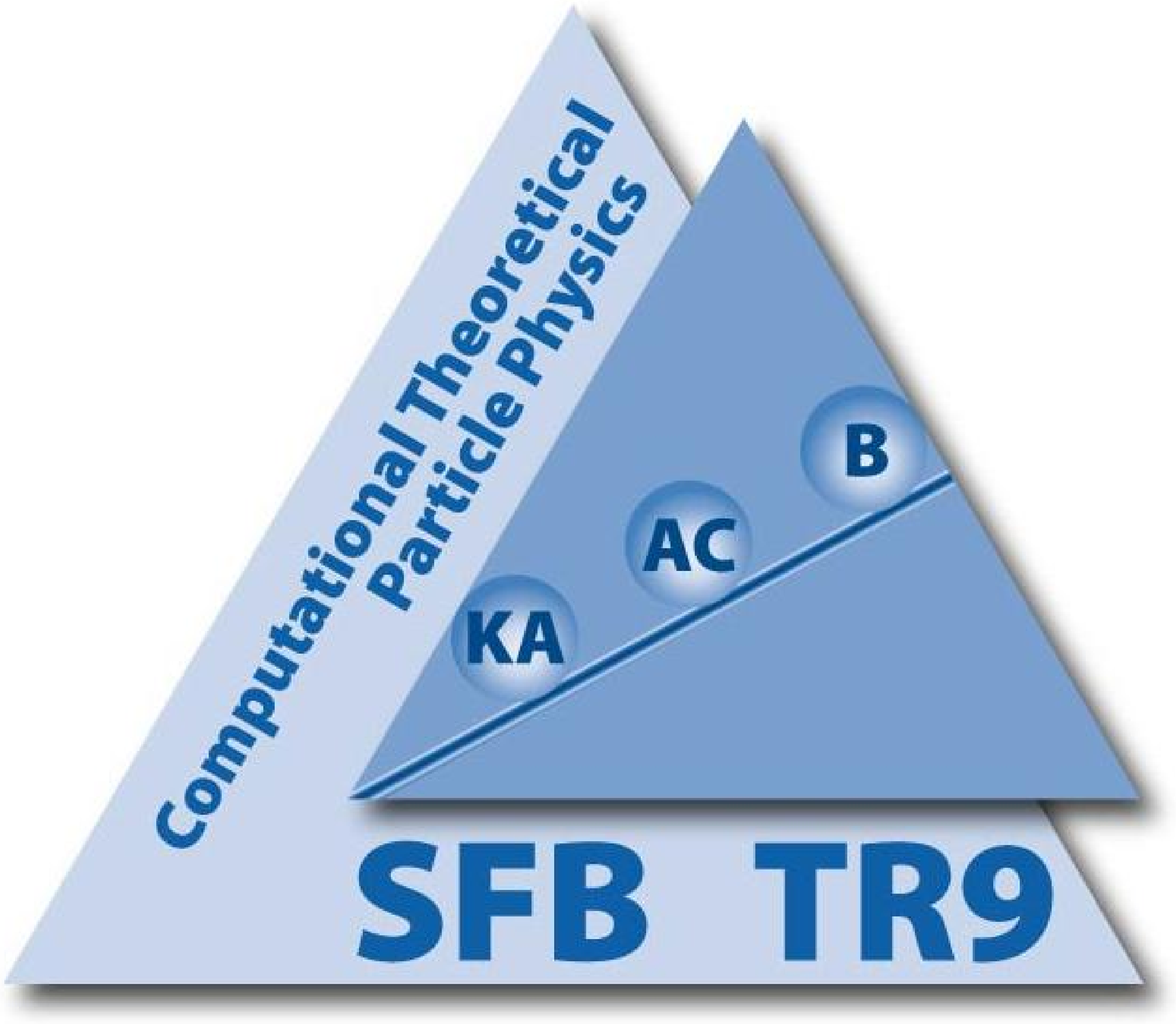} \\
          {\it \hfill Alberta-Thy-14-07\\
          \hfill DESY 07-165 \\ 
          \hfill MIT-CTP 3874\\
          \hfill SFB/CPP-07-57\\
          \hfill TTP07-26}\\
          $ $\vspace{-4.5cm}\\
        \llap{$^a$} Department of Physics, University of Alberta\\
                    Edmonton, Alberta T6G 2G7, Canada\\
                    \\
        \llap{$^b$} Budker Institute of Nuclear Physics\\
                    Novosibirsk 630090, Russia\\
                    \\
        \llap{$^c$} Deutsches Elektronen-Synchrotron (DESY)\\
                    Platanenallee 6, 15738 Zeuthen, Germany\\
                    \\
        \llap{$^d$} Institut \fuer Theoretische Teilchenphysik\\
                    \DUNI Karlsruhe, 76128 Karlsruhe, Germany\\ 
                    \\
        \llap{$^e$} Center for Theoretical Physics, Massachusetts Institute of Technology \\
                    Cambridge, MA 02139, U.S.A.\\

          \\
        E-mail: \email{a.g.grozin@inp.nsk.su}, \email{damiano.guazzini@desy.de}, 
                \email{peter.marquard@kit.edu}, \email{meyerh@mit.edu}, \email{jpiclum@phys.ualberta.ca},
                 \email{rainer.sommer@desy.de}, \email{matthias.steinhauser@uka.de},}

\abstract{We perform the non-perturbative (quenched) renormalization of the chromo-magnetic operator in Heavy Quark 
          Effective Theory and its three-loop matching to QCD. At order $1/m$ of the expansion, the operator is responsible
          for the mass splitting between the pseudoscalar and vector B-mesons. 
          These new computed factors are affected by an uncertainty negligible in comparison to  
          the known bare matrix element of the operator between B-states. 
          Furthermore, they push the quenched determination of the 
          spin splitting for the ${\rm B_s}$-meson much closer to its experimental value 
          than the previous perturbatively renormalized
          computations. The renormalization factor for three commonly used
          heavy quark actions and the Wilson gauge action and useful parametrizations of the matching
          coefficient are provided.}

\FullConference{The XXV International Symposium on Lattice Field Theory\\
		 July 30-4 August 2007\\
		 Regensburg, Germany}

\begin{document}

\section{The effective theory and the chromo-magnetic operator}\label{sec:HQET}

We consider the classical HQET Lagrangian \cite{Eichten:1989zv,Grinstein:1990mj,Georgi:1990um} of a heavy fermion of 
mass\footnote{The details upon the heavy quark mass definition are irrelevant for the present discussion.} $m$, 
whose spinor we indicate with $\heavy$.
Keeping a four component notation with $P_+\heavy=\heavy$ we thus have
\bea
{\cal L}&=&{\cal L}^{\rm stat}+{\cal L}^{(1)}+{\rm O}(1/m^2)\,,\label{e:Lagr}\\
{\cal L}^{\rm stat}&=&\heavyb D_0 \heavy\,,\qquad {\cal L}^{(1)}=-\frac{1}{2m}({\cal O}_{\rm kin}+{\cal O}_{\rm spin})
=\frac{1}{2m}\heavyb (-\overrightarrow{D}\phantom{D}^{\hspace{-0.28cm}2}
-\frac{1}{2i}F_{kl}\sigma_{kl})\heavy\,,\\
{\cal O}_{\rm kin}&=&\heavyb \overrightarrow{D}\phantom{D}^{\hspace{-0.28cm}2} \heavy\,,\qquad{\cal O}_{\rm spin}=\heavyb\frac{1}{2i}F_{kl}\sigma_{kl} \heavy=
\heavyb\overrightarrow{\sigma}\!\cdot\! \overrightarrow{B}\heavy\,.
\eea
where $\overrightarrow{D}\phantom{D}^{\hspace{-0.28cm}2}=D_kD_k$, $\sigma_{kl}=\frac{i}{2}[\gamma_k,\gamma_l]$ 
and $F_{\mu\nu}$ is the QCD field strength tensor.
The spin-flavor symmetry of the static Lagrangian ${\cal L}^{\rm stat}$
is broken at the ${\rm O}(1/m)$ by the kinetic and the chromo-magnetic operators.  
At this order only the latter is responsible for 
the spin interaction. In particular
the quadratic mass splitting between the ground state
pseudoscalar (PS) and vector (V) heavy-light mesons assumes the form
\be\label{e:ssp1}
\Delta m^2=M_{\rm V}^2-M_{\rm PS}^2=4\lambda_2+{\rm O}(\Lambda_{\rm QCD}^3/m)\,.
\ee
The parameter $\lambda_2$ is directly related to $\Ospin$ and encodes, at order $1/m$, the information upon 
the deviations from the static limit, where $M_{\rm V}=M_{\rm PS}$,
stemming from the spin-dependent interactions inside the heavy-light mesons. 
The splitting (\ref{e:ssp1}) can be
rewritten in two equivalent ways
\be\label{e:Delta_m_RGI}
\Delta m^2=4 C_{\rm mag}(M/\Lambda)\lambda_2^{\rm RGI}+{\rm O}(\frac{\Lambda^3}{m})=
2{M_{\rm V}+M_{\rm PS}\over M}C_{\rm spin}(M/\Lambda)\lambda_2^{\rm RGI}+{\rm O}(\frac{\Lambda^3}{m})\,,\quad 
\Lambda=\Lambda_{\MSbar}\,.
\ee
The coefficients $C_{\rm mag}$ and $C_{\rm spin}$ perform the matching between HQET and QCD,
and are expressed as functions of the RGI heavy quark mass $M$, defined as in \cite{Capitani:1998mq}. They are computable
in continuum perturbation theory, and a three-loop result is presented in \sect{sec:3loop}, where 
a motivation for preferring the second form in (\ref{e:Delta_m_RGI}) is provided. The RGI parameter $\lambda_2^{\rm RGI}$
is given by
\bea
\lambda_2^{\rm RGI}&=&\frac{1}{3}\langle {\rm B}|{\cal O}_{\rm spin}^{\rm RGI} |{\rm B} \rangle/
\langle {\rm B}|{\rm B}\rangle\,,\qquad
{\cal O}_{\rm spin}^{\rm RGI}=\lim_{\mu\to \infty}
[2b_0\gbsq(\mu)]^{-\gamma_0/2b_0}{\cal O}^S_{\rm spin}(\mu)\,,\label{e:lambda2_RGI}\\
\mbox{with}\qquad\gamma_0&=&3/(8\pi^2)\,,\qquad b_0=(11-\frac{2}{3}N_{\rm f})/(16\pi^2)\,,\label{e:g0b0}
\eea
and the zero-momentum static-light meson state $|{\rm B}\rangle$. The operator 
${\cal O}^S_{\rm spin}(\mu)$ is related to the bare operator ${\cal O}_{\rm spin}$ by a multiplicative renormalization
factor $Z_{\rm spin}^S(\mu)$ depending on the adopted scheme $S$ and a renormalization scale $\mu$, whereas
$Z_{\rm spin}^{\rm RGI}(g_0)={\cal O}_{\rm spin}^{\rm RGI}/{\cal O}_{\rm spin}$ depends on the bare coupling only. The relation
between the two renormalization factors reads 
\be\label{e:ZRGIoverZS}
   Z_{\rm spin}^S(\mu)/Z_{\rm spin}^{\rm RGI}=\Phi_{\rm spin}^{S}(\mu)/\Phi_{\rm spin}^{\rm RGI}=U^S(\mu)\,,
\ee
where 
\be \label{e:U}
U^S(\mu) =[2b_0\bar{g}^2_S(\mu)]^{\gamma_0/2b_0}{\rm exp}\left\{\int_0^{\bar{g}_S(\mu)}
{\rm d}g \left[\frac{\gamma^S(g)}{\beta^S(g)}-\frac{\gamma_0}{b_0 g}\right]\right\}\,,
\ee
is the solution of the renormalization group equation in terms of
the anomalous dimension $\gamma^S$ and the $\beta$-function in the $S$ scheme with their leading
order coupling expansion coefficients (\ref{e:g0b0}). Here $\Phi$ stands for any
matrix element of $\Ospin$, e.g.~$\lambda_2$.

\section{Non-perturbative renormalization}\label{sec:NPren}

We follow the general strategy of \cite{Capitani:1998mq}, and formulate a renormalization condition
for ${\cal O}_{\rm spin}$ in a finite volume, which enables us to non-perturbatively compute the 
renormalization factor $Z_{\rm spin}^{\rm RGI}$. As we are interested in accurate simulations as well as
perturbative computations we choose \SF (SF) boundary conditions; see \cite{Sommer:2006sj} for a recent review. 
They induce a non-trivial background field, $F_{\mu\nu}$, at tree-level.
This ensures a good signal in MC simulations at weak coupling. Further, it means that a 1-loop computation
is sufficient to know the renormalization factor up to and including ${\rm O}(g_0^2)$. 
Since $\Ospin$ does not contain any light fermion fields, we are able to avoid these altogether in the definition
of the correlation functions. It follows that for $N_{\rm f}=0$
we end up with a pure gauge theory definition (with no relativistic valence quarks) and the
observables are $\Oa$-improved, once the action is. 

In a discretized box of volume $L^4$ we adopt Dirichlet boundary
conditions in the $\hat 3$-direction and periodic boundary conditions in all others. 
A natural renormalization condition is then 
\be\label{e:ren_cond}
Z_{\rm spin}^{\rm SF}(L){L^2 \langle S_1(x+\frac{L}{2}\hat 0)\Ospin(x)\rangle \over \langle S_1(x+\frac{L}{2}\hat 0)S_1(x)\rangle}=
\left.{L^2 \langle S_1(x+\frac{L}{2}\hat 0)\Ospin(x)\rangle\over \langle S_1(x+\frac{L}{2}\hat 0)S_1(x)\rangle}\right|_{g_0=0}\,,
\quad x_3=L/2\,.
\ee
The spin operator
$S_1(x)={1\over 1+a\delta m_{\rm W}}\heavyb\sigma_1{W}^{\dagger}_0 (x-a\hat 0)\heavy(x-a\hat 0)$
is introduced in order to have a non-vanishing trace in spin space. It is a (local) Noether charge and does not
need to be renormalized. ${\rm W}_0$ is the same temporal parallel transporter appearing in the 
discretized static action \cite{Della_Morte:2005yc}, 
and $\delta m_{\rm W}$
is an additive mass renormalization term, whose knowledge is not needed in the following; it cancels out in the
ratios of \eq{e:ren_cond}. 

After integrating the static quark fields out and exploiting the properties 
of the static propagator \cite{Kurth:2000ki,Della_Morte:2005yc}, we use the equivalence
of all coordinates in Euclidean space to switch to the usual SF boundary conditions, 
corresponding to ``point A'' in  \cite{Luscher:1993gh}, and obtain
\be\label{e:ren_cond2} 
Z_{\rm spin}^{\rm SF}(L) {L^2\langle  \Tr({\cal P}_3(x)E_1(x)) \rangle \over \langle \Tr({\cal P}_3(x)) \rangle} 
=\left. {L^2\langle  \Tr({\cal P}_3(x)E_1(x)) \rangle \over \langle \Tr({\cal P}_3(x)) \rangle} \right|_{g_0=0}
={\pi\over 6}{1+\sqrt{3}\over 2-\sqrt{3}}+{\rm O}((a/L)^4)\,,
\ee
with $x_0=L_0/2$, $E_1=i\hat{F}_{01}(x)$, and Dirichlet boundary conditions in time.
Here $\hat{F}_{01}(x)$ stands for the clover leaf discretization of the field strength tensor \cite{Luscher:1996sc}.

\begin{figure}[t]
\vspace{-0.4cm}
\begin{center}
\begin{tabular}{cc}
\includegraphics[height=5.5cm]{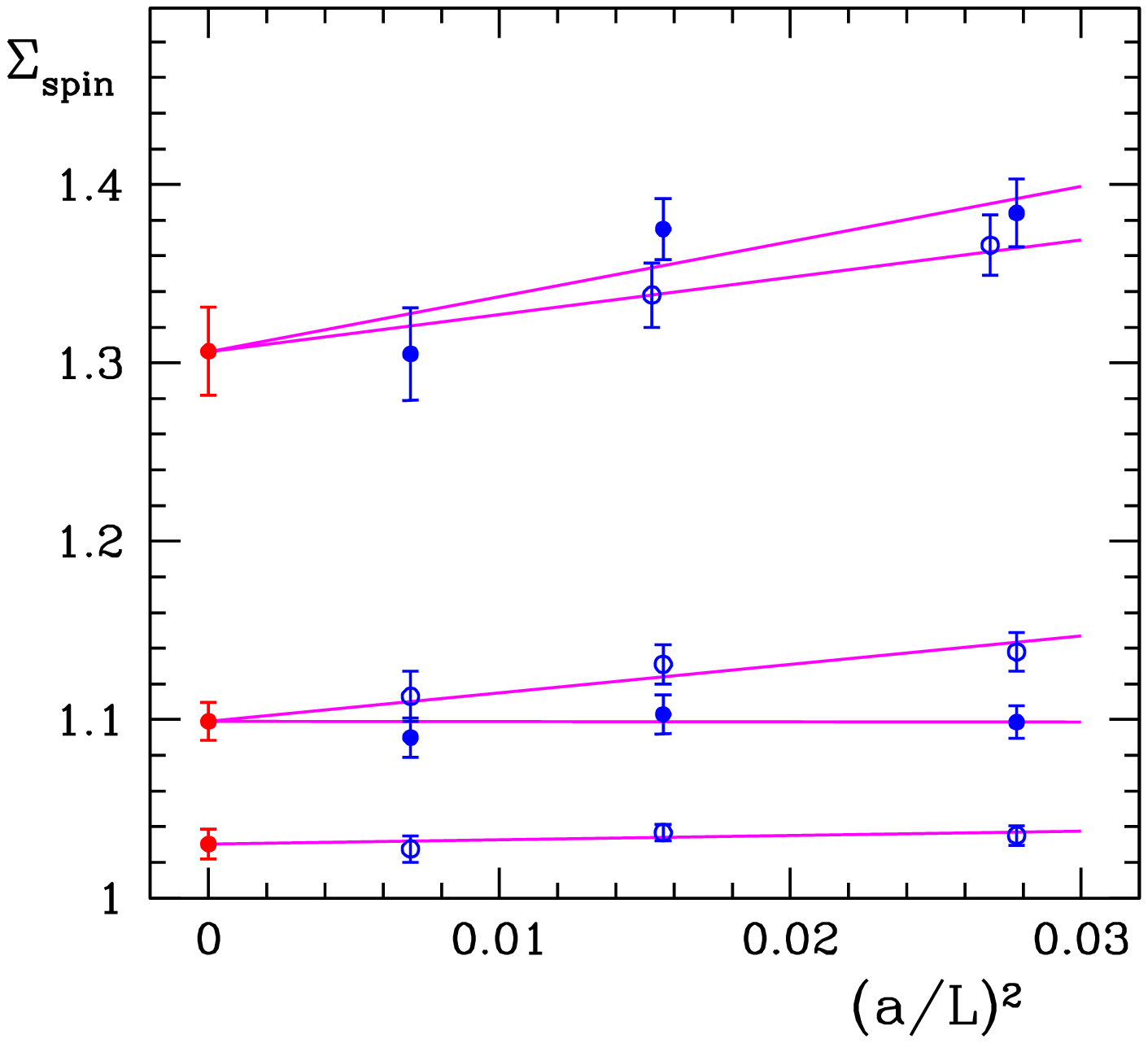} &
\includegraphics[height=5.5cm]{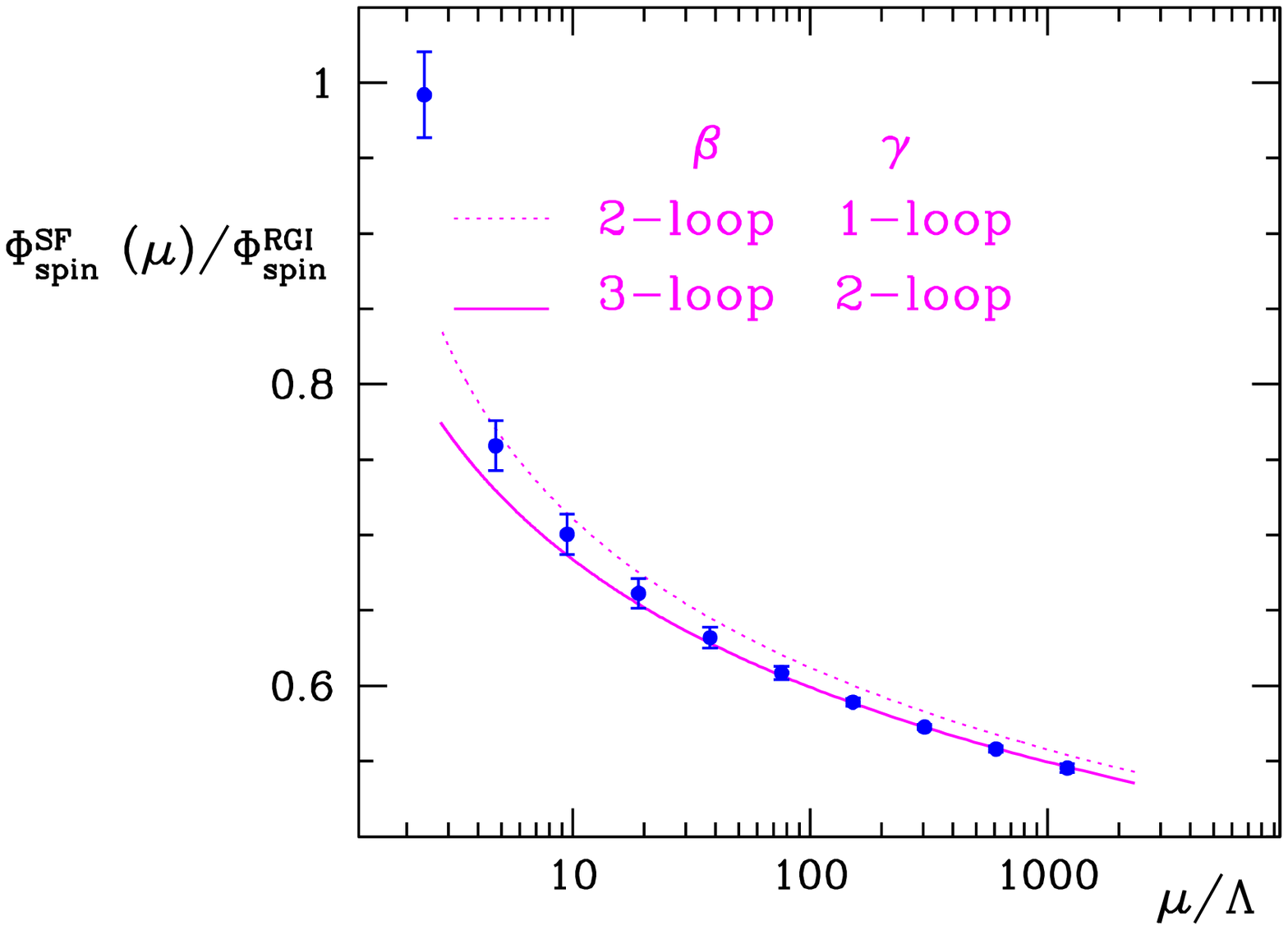} \qquad \\
\end{tabular}
\end{center}
\vspace{-0.5cm}
\caption[]{\label{fig:Sigma_ospin}
{Left: examples of continuum limit extrapolations of $\Sigma_{ \rm spin}$ (cf.~\eqs{e:S_spin}) for
couplings $u=1.243, 2.77\,\mbox{and}\,3.48$. Filled symbols indicate that $F_{\mu\nu}$ was defined as $\hat{F}_{\mu\nu}$
with the link variables replaced by HYP2 \cite{Della_Morte:2005yc} links. Right: scale dependence of $\Ospin$ in the SF scheme with
its associated 
$\Lambda$ parameter~\cite{Luscher:1993gh}.}}
\end{figure}

Having specified the lattice setup and the renormalization condition, we introduce 
the step scaling function $\sigma_{\rm spin}(u)$ via
\be\label{e:sigma}
\Ospin^{\rm SF}(\mu)=\sigma_{\rm spin}(\gbsq(1/\mu))\Ospin^{\rm SF}(2\mu)\,.
\ee
It is obtained as the continuum limit\\ [-3ex]
\be\label{e:S_spin}
\sigma_{\rm spin}(u)=\lim_{a/L\to 0}\Sigma_{\rm spin}(u,a/L)\quad\mbox{of}\quad
\Sigma_{\rm spin}(u,a/L)=\left. {Z_{\rm spin}^{\rm SF}(2L)\over Z_{\rm spin}^{\rm SF}(L)}\right|_{\gbsq(L)=u\,,\, m=0}\,,
\ee
where $\gbsq(L)$ is the SF coupling and the condition $m=0$ of vanishing
light quark masses plays a role only in case that the computation is extended to $N_{\rm f}>0$.
We performed pure gauge theory simulations to determine $\Sigma_{\rm spin}$ for different couplings $u$ and
resolutions $a/L$. The continuum limit results (see \Fig{fig:Sigma_ospin}) allow us 
to reconstruct the non-perturbative scale dependence
of the SF renormalized chromo-magnetic operator.

By applying eqs.~$(\ref{e:g0b0},\, \ref{e:ZRGIoverZS})$ at weak coupling $\gbsq(\mu)$ with the two-loop anomalous dimension in the 
SF scheme 
\cite{Guazzini:2007bu,Damiano:thesis},
\be\label{e:gamma_1_SF}
\gamma^{\rm SF}(\bar{g})=-\gbsq(\gamma_0+\gamma_1^{\rm SF}\gbsq+\ldots)\,,\quad
\gamma_1^{\rm SF}=-0.00236-0.00352 N_{\rm f}+0.00023N_{\rm f}^2\,,
\ee
we are able to non-perturbatively connect the low energy regime with the RGI, 
and arrive at 
\bea
\Phi^{\rm SF}_{\rm spin}(\mu)/\Phi^{\rm RGI}_{\rm spin}&=&0.992(29)\,,\qquad\mbox{at $\mu=1/2L_{\rm max}$}\,,\label{e:Phi_2Lmax}
\qquad 2L_{\rm max}=1.436 r_0 \,\cite{Guagnelli:1998ud}\,.
\eea
The latter has to be combined with values of $Z_{\rm spin}^{\rm SF}(2L_{\rm max})$, depending
on the bare coupling and lattice action, to form
\be\label{e:ZspinRGI}
Z_{\rm spin}^{\rm RGI}=Z_{\rm spin}^{\rm SF}(L)× \Phi^{\rm RGI}_{\rm spin}/ \Phi^{\rm SF}_{\rm spin}(1/L)
\ee
for the respective action. The numerical values are well represented by
\be
Z_{\rm spin}^{\rm SF}(2L_{\rm max})=2.55+0.16(\beta-6)-0.40(\beta-6)^2\,, \qquad 6.0\leq \beta\leq 6.5\,,
\ee
for the HYP1 \cite{Della_Morte:2005yc} action with an error of about 1\%. For the other actions see \cite{Guazzini:2007bu}.

\section{Three-loop matching between HQET and QCD}\label{sec:3loop}

As pointed out in \sect{sec:HQET} the perturbative matching between HQET and QCD
plays a very important role in a precise determination of the mass splitting.
Our three-loop computation \cite{Grozin:2007fh} of the matching coefficient
and the anomalous dimension of the
chromo-magnetic operator allow us to give a reliable final result and
estimate its uncertainty.

The coefficient of the chromo-magnetic term needs to be determined by matching to QCD. 
In perturbation theory we consider the scattering amplitude
of an on-shell heavy quark in an external chromo-magnetic field, expanded in the momentum transfer $q$ up to
the linear term. Denoting it schematically  by $\ampl$ and indicating only
the presently relevant dependences, we have the (traditional) matching condition 
(with $\amplmsbar(\mu)=U^{\msbar}(\mu)\, \amplhqet$ as in \eq{e:U})
\be
  {\amplqcd} = {1 \over m_{\rm Q}}\, C_{\rm cm}(m_{\rm Q}) U^{\msbar}(m_{\rm Q})\, \amplhqet \,,\quad 
  \amplhqet = \langle \beta | \Ospinrgi | \alpha \rangle \,.
\ee
By working in the $\MSbar$ scheme and with the background field
method \cite{Abbott:1980hw}, we arrive at the 3-loop result for the
matching coefficient
\bea
C_{\rm cm}(m_{\rm Q})&=&1+0.6897\alphaMSbar(m_{\rm
Q})+(2.2182-0.1938N_{\rm f})\alphaMSbar^2(m_{\rm Q})\nonumber\\[-1.8ex]
& &\label{e:Ccm}\\[-1.8ex]
& &+(11.0763-1.7495N_{\rm f}+0.0513N_{\rm f}^2)\alphaMSbar^3(m_{\rm
Q})+{\rm O}(\alphaMSbar^4)\,,\nonumber
\eea
while for the anomalous dimension of $\Ospin^\msbar$,
which enters $U^\msbar(\mu)$, we extract
\bea
\gamma^{\MSbar}(\alphaMSbar)&=&0.4775 \alphaMSbar+(0.4306-0.0549N_{\rm
f})\alphaMSbar^2\nonumber\\[-1.8ex]
& &\label{e:gamma_cm}\\[-1.8ex]
& &+(0.8823-0.1472N_{\rm f}-0.0007N_{\rm f}^2)\alphaMSbar^3+{\rm
O}(\alphaMSbar^4)\,.\nonumber
\eea
Here, formulae are given for the case where the heavy quarks are quenched
also in QCD. Their loop effects are {\em very} small \cite{Grozin:2007fh}.
The conversion function $\Cmag$ of \sect{sec:HQET} is obtained by changing
the renormalization scheme in the effective theory such as to include 
the finite renormalization $C_{\rm cm}$, while $\Cspin$ is constructed by
replacing in addition the pole mass, $m_{\rm Q}$, by the RGI mass, $M$:
\be \label{e:magspin}
  \amplqcd =  {1 \over m_{\rm Q}}\,\Cmag(M/\Lambda_\msbar)\, \amplhqet 
  =  {1 \over M}\,\Cspin(M/\Lambda_\msbar)\, \amplhqet \,.
\ee
The resulting equations 
\be
  \Cspin(M/\Lambda_\msbar) \equiv U^\mrm{spin}(\mbar_*)
  = {M \over m_{\rm Q}} \Cmag(M/\Lambda_\msbar) \equiv   {M \over m_{\rm Q}} U^\mrm{mag}(\mbar_*)
\ee
then define the anomalous dimensions $\gamma^\mrm{spin}\,,\,\gamma^\mrm{mag}$. In all these 
schemes the renormalization of the coupling remains untouched: $\msbar$. The change
from the $\msbar$-mass at its own scale $\mbar_*$ as argument of $U^\mrm{spin}$ 
to the RGI-mass as the argument of 
$\Cspin$ is convenient since the RGI-masses are the primary quantities obtained
in a non-perturbative lattice computation \cite{Capitani:1998mq}. 

The second equation
in (\ref{e:magspin}) avoids the pole mass which is known to have a bad perturbative
expansion in terms of short distance masses (or $M$). Thus the anomalous
dimension $\gamma^\mrm{spin}$ is expected
to show a better behaved 
perturbative series which will be reflected in $\Cspin$.

\begin{figure}[t]
\vspace{-0.4cm}
\begin{center}
\includegraphics[scale=0.6]{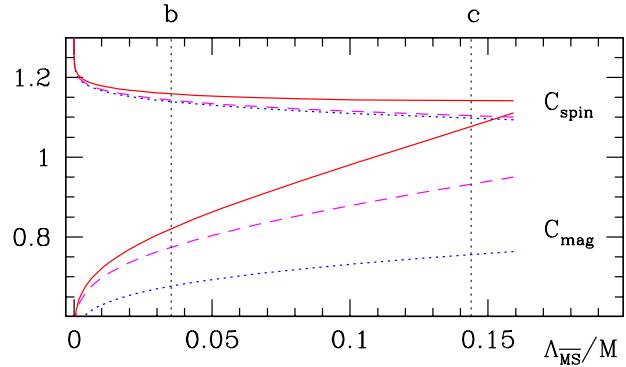} 
\end{center}
\vspace{-0.5cm}
\caption[]{\label{fig:C_spin3}
{Conversion functions for $N_{\rm f}=0$. Dotted, dashed and solid lines use the one-, two- and three-loop anomalous
dimension. The abscissae of the b- and 
c-quark \cite{Della_Morte:2006cb,Damiano:thesis,Guazzini:2006bn,Rolf:2002gu} 
are marked by dotted lines.}}
\end{figure}

For practical purposes we parametrize the conversion functions $C_{\rm spin}$ and $C_{\rm mag}$
in the $\nf=0$ theory, graphically
represented in \Fig{fig:C_spin3}, in terms
of the variable $x\equiv 1/\ln(M/\Lambda_\msbar)$:
\be
C_{\rm spin}=\left\{ \begin{array}{ll}
x^{\gamma_0^{\rm spin}/(2b_0)}\{1+0.087x-0.021x^2\}&\mbox{2-loop $\gamma$}\\
x^{\gamma_0^{\rm spin}/(2b_0)}\{1+0.097x+0.115x^2-0.038x^3\}&\mbox{3-loop $\gamma$}\label{e:Cspin_para}\\
\end{array}\right.\,,\quad\gamma_0^{\rm spin}=-2/(4\pi)^2\,.
\ee
These formulae guarantee at least 0.3\% precision for $x\leq 0.6$. Inspection of 
\Fig{fig:C_spin3} shows the expected 
bad perturbative behavior of $C_{\rm mag}$. We thus 
focus our attention on $C_{\rm spin}$ which exhibits very small higher order contributions 
in the b-region.
The difference $\Delta C_{\rm spin}(M_{\rm b}/\Lambda)\approx 10^{-2}$ between the 
three-loop and the two-loop determination
with $M_{\rm b}=6.76(9)\,\GeV$ (from \cite{Della_Morte:2006cb})
is much smaller than the statistical error on the spin splitting presented in the following section. 
Evaluating it with an estimate (where the four-loop term in the very well behaved
$\gamma^\msbar$ is neglected) for the anomalous dimension
$\gamma^\mrm{spin}$ gives
$\Delta C_{\rm spin}(M_{\rm b}/\Lambda)\approx 10^{-2}$ with respect to the three-loop estimate. We thus
claim an about 1\% relative error for $C_{\rm spin}$ evaluated with the three-loop $\gamma^\mrm{spin}$
for B-physics applications. For $N_{\rm f}=4$ the behavior of
$C_{\rm mag}$ and $C_{\rm spin}$ is very similar to \fig{fig:C_spin3}~\cite{Grozin:2007fh}. 

\section{First results for the spin-splitting and outlook}\label{sec:spsp}

As a first application we take quenched results for the bare $\lambda_2$ from the literature
and exploit our results (\ref{e:ZspinRGI}, \ref{e:Cspin_para}). Unfortunately they exist only for $\beta=6.0$,
corresponding to $a\approx 0.1$ fm, 
\bea
& &\mbox{Ref.}~\cite{Gimenez:1996av}:\quad \Delta m^2=0.28(6)(?)\,\GeV^2\,\, 
\stackrel{(\ref{e:ZspinRGI},\,\ref{e:Cspin_para})}{\longrightarrow}\,\,\Delta m^2=0.38(7)(?)\,\GeV^2\,,\\
& &\mbox{Ref.}~\cite{Aoki:2003jf}:\quad \Delta m^2=0.36(4)(?)\,\GeV^2\,\, 
\stackrel{(\ref{e:ZspinRGI},\, \ref{e:Cspin_para})}{\longrightarrow}\,\,\Delta m^2=0.53(6)(?)\,\GeV^2\,,
\eea 
where the numbers on the l.h.s.~are taken from the corresponding references, performing a perturbative renormalization.
On the r.h.s.~we used the b-quark mass from \cite{Della_Morte:2006cb} and the 3-loop determination of $C_{\rm spin}$.
The uncertainty marked as $(?)$ refers to lattice artefacts and the missing dynamical quark determinant. 
The central values are now closer to the experimental mass splitting, $\Delta m^2=0.497\,\GeV^2$, but at the moment
the large
uncertainties prevent us from concluding that indeed the quenched approximation can give a good estimate of this 
observable.

As explained in \cite{Guazzini:2007bu}, the same renormalization factor applies 
to spin-dependent potentials \cite{Eichten:1980mw,Vairo:2007id},
where so far only a perturbative renormalization was possible. 
 
The non-perturbative computation of $Z_{\rm spin}^{\rm RGI}$ has demonstrated the
applicability of the \SF renormalization programme \cite{Luscher:1991wu,Bode:2001jv}
to  another difficult case. 
Quite significant deviations from the perturbative scale evolution
are present at low energies, see \Fig{fig:Sigma_ospin}. 
 
With respect to a perturbative estimate, the new $Z_{\rm spin}^{\rm RGI}$ has a rather big effect.
Furthermore, thanks to the results presented in \sect{sec:3loop}, which extend 
\cite{Eichten:1990vp,Falk:1990pz,Amoros:1997rx,Czarnecki:1997dz,Heitger:2004gb}, we can match the 
effective theory and QCD introducing
an error in practice negligible in comparison to all other uncertainties entering $\Delta m^2$.
It now remains to compute $\lambda_2^{\rm bare}$ with higher precision and perform the continuum 
limit. However, due to the large amount of statistics needed
especially at large couplings, an extension of this method to the dynamical quarks case seems difficult. 
In this direction, other, fully non-perturbative, approaches are more promising at present
\cite{Guazzini:2006bn, Damiano:thesis, Heitger:2003nj, Della_Morte:2006cb}.\\[-2ex]

\noindent {\bf Acknowledgements.}
We thank M.~Della Morte, J.~Flynn, B.~Leder, S.~Takeda and U.~Wolff for fruitful discussions.  
This work is supported by the  Deutsche Forschungsgemeinschaft 
in the SFB/TR~09, by the European community through 
EU Contract No.~MRTN-CT-2006-035482, ``FLAVIAnet''
and by funds provided by the U.S. Department of Energy under 
cooperative research agreement DE-FC02-94ER40818.

\bibliographystyle{h-elsevier}
\bibliography{refs}

\end{document}